\documentclass[11pt,aps,preprint]{revtex4}
\usepackage{graphicx}
\usepackage{amsmath}
\usepackage{amsfonts}
\usepackage{amssymb}
\begin{document}
\title{Theory of subgap interchain tunneling in quasi 1D conductors}
\author{S. Brazovskii}
\affiliation{LPTMS-CNRS, UMR 8626, Univ. Paris-Sud, bat. 100, Orsay, 91405, France}
\author{S.I. Matveenko}
 \affiliation{L.D. Landau Institute for Theoretical Physics,
  Kosygina Str. 2, 119334, Moscow, Russia}

\date{10/09/2007}

\begin{abstract}
We suggest a theory of internal coherent tunneling in the pseudogap region, when the applied voltage $U$ is below the free electron gap $2\Delta_0$. We address quasi 1D systems, where the gap is originated by spontaneous lattice distortions of the Incommensurate Charge Density Wave (ICDW) type. Results can be adjusted also to quasi-1D superconductors. The instanton approach allows to calculate the interchain tunneling current both in single electron (amplitude solitons, i.e. spinons) and bi-electron (phase slips) channels. Transition rates are governed by a dissipative dynamics originated by emission of gapless phase excitations in the course of the instanton process. We find that the single-electron tunneling is allowed down to the true pair-breaking threshold at $U_c=2W_{as}<2\Delta_{0}$, where $W_{as}=2/\pi\Delta_{0}$ is the amplitude soliton energy. Most importantly, the bi-electronic tunneling stretches down to $U_c=0$ (in the 1D regime). In both cases, the threshold behavior is given by power laws $J\sim(U-U_c)^{\beta}$, where the exponent ${\beta}\sim v_F/u$ is large as the ratio of the Fermi velocity $v_F$ and the phase one $u$. In the 2D or 3D ordered phases, at temperature $T<T_c$, the one-electron tunneling current does not vanish at the threshold anymore, but saturates above it at $U-U_c\sim T_c\ll\Delta_0$. Also the bi-particle channel acquires a finite threshold $U_{c}=W_{2\pi}\sim T_c\ll\Delta_0$ at the energy of the $2\pi$
phase soliton.
\end{abstract}

\maketitle

\section{Introduction.}

\subsection{Pseudogaps and the subgap tunneling.}

Interchain, interplane transport of electrons in low dimensional (quasi 1D,2D) materials attracts much attention in view of striking differences between longitudinal and transverse transport mechanisms, revealing a general problematics of strongly correlated electronic systems \cite{pwa:98}. Beyond a low field (linear) conduction, the whole tunneling current-voltage J-U characteristic $J(U)$ and the conductivity $\sigma=dJ/dU$ are of particular importance. A common interest in tunneling phenomena in quasi 1D conductors has been further endorsed by recent experimental achievements, see \cite{zandt}, and especially in new techniques of the intrinsic tunneling (see \cite{latyshev:05-ecrys} for a short review) where electronic interchain transitions take place in the bulk of the unperturbed material. Particularly appealing is the access to topologically nontrivial excitations - solitons, showing up in subgap  spectra \cite{latyshev-prl:05,latyshev-prl:06,brazov:ecrys-05}. Role of preexisting solitons in tunneling spectra of High-$\mathrm{T_c}$ materials has been advocated in \cite{mourachkine}. The goal of this article is to describe their manifestations and peculiarities in Incommensurate Charge Density Wave (ICDW), see \cite{gorkov:89,ECRYS02,ECRYS05}. The following content of this chapter will introduce experimental evidences, theoretical grounds and conditions for existence of solitons. Chapter II will describe the theoretical approach. Chapter III will give results of calculations, which details can be found in the Appendix.

If the tunneling processes were going between free electron states, as they have been formed by the rigid ICDW, then the current onset would correspond to the voltage $E_{g}^{0}=2\Delta_{0}$ of the gap in the spectrum of electrons. But actually there is also a possibility for tunneling within the subgap region $E_{g}<U<E_{g}^{0}$. It is related to the pseudogap (PG) phenomenon known for strongly correlated electrons in general, well pronounced in quasi 1D systems, and particularly in cases where the gap is opened by a spontaneous symmetry breaking (see \cite{matv-ecrys02,matveenko:05-ecrys,brazov:03,mb:02,mb:05} and refs. therein). The PG is originated by a difference, sometimes qualitative, between short living excitations which are close to free electrons, and dressed stationary excitations of the whole correlated systems. The true excitations gain the energy, but loose the probability of transitions to these complex states. In the ICDW, the dressing of a bare single particle results in self-trapped states like the amplitude solitons with energies $W_{as}<\Delta_{0}$ below the free electron activation energy $\Delta_{0}$; then the subgap interval $2W_{as}<U<2\Delta_{0}$ will be observed as a PG. Below this pair-breaking threshold $2W_{as}$ (observed in optics, spin susceptibility), in tunneling there may be also contributions of collective states - phase solitons, which energy $W_{ps}$ is even much lower. These states are activated in the bi-electronic channel - coherent tunneling of two electrons. In the 1D regime (which we shall mostly address below) $W_{ps}=0$, then the subgap tunneling fils the whole pseudogap.

\subsection{Solitons in ICDWs}

Strongly correlated electronic systems show various types of symmetry breaking originating degenerate ground states (GS). The degeneracy gives rise to topologically nontrivial perturbations exploring the possibility of traveling among different allowed GSs. Of special interest are totally localized and truly microscopic objects, namely solitons (or instantons for related transient processes) which can carry single electronic quantum numbers: either spin $1/2$ or charge $e$, or both. Being energetically favorable with respect to electrons, solitons would determine the electronic properties which can be proved theoretically at least for quasi 1D systems. We address here the cases of a continuous GS degeneracy of an Incommensurate CDW (ICDW) (see \cite{gorkov:89,ECRYS02,ECRYS05}) where solitons can be created in single items via single electronic processes. Results can be generalized also to quasi-1D superconductors. Continuous degeneracy of the ICDW order parameter $A\cos(Qx+\varphi)$ comes from an arbitrary chosen phase $\varphi$, which characterizes the freedom of translation $\delta x=-\delta\varphi/Q$ of the ICDW as a whole. At a first sight, the CDW is commonly viewed as a narrow gap anisotropic semiconductor with most of its properties described by free electrons $e$ or holes $h$ near the gap edges $\pm\Delta_{0}$. Thus, $\Delta_{0}$ would give the activation energy in kinetics and thermodynamics characteristics (conductivity, spin susceptibility, heat capacitance, NMR); the same $\Delta_{0}$ would be observed from dynamic probes as photoemission, external tunneling. The double gap $2\Delta_{0}$ would be measured as the edge in optics or in internal tunneling described below. Doping or junction injection (FET) would require the threshold $\Delta_{0}$ and lead to formation of electronic pockets near the gap edges. But actually \emph{almost nothing from this well established semiconducting picture} takes place in ICDWs.
\newline1. Activation energies ($\Delta_{\parallel}$ and $\Delta_{\perp}$)
measured from transport in the on-chain ($\Delta_{\parallel}$) and
in the interchain ($\Delta_{\perp}$) directions differ by several
times (typical values for TaS$_{3}$ are
$\Delta_{\parallel}\approx200$~K and $\Delta_{\perp}\approx800$~K
\cite{nad:89}) which signifies an intrinsically different character of processes involved;
\newline2. Energies deduced from the spin activation $\Delta_{s}~=640~K$ \cite{johnston:83} and from the dynamic
relaxation $\Delta _{d}$ \cite{demsar:99} are in between $\approx600~$K;
\newline3. Optical absorption shows the peak at the
scale of $2\Delta_{\perp}$ - thus associated to $2\Delta_{0}$, but
its shape is strongly smeared out with intensity deeply spread within the expected spectral gap \cite{degiorgi:95} - the pseudogap effect;
\newline4. Thresholds for charge transfer may not exist at
all, or have low values compatible with $\Delta_{\parallel}$ - i.e.
associated with the interchain decoupling scale, see
\cite{latyshev:05-ecrys};
\newline5. Charge injection results in absorption of additional electrons into the extended GS via
so-called phase slip processes (see \cite{gorkov:89a}), rather than in formation of Fermi pockets. The static phase slip - a $2\pi$ soliton as an on-chain CDW defect, has been directly observed in recent STM experiments \cite{brun}.

\begin{figure}[tbh]
\begin{center}
\includegraphics[width=0.5\textwidth]{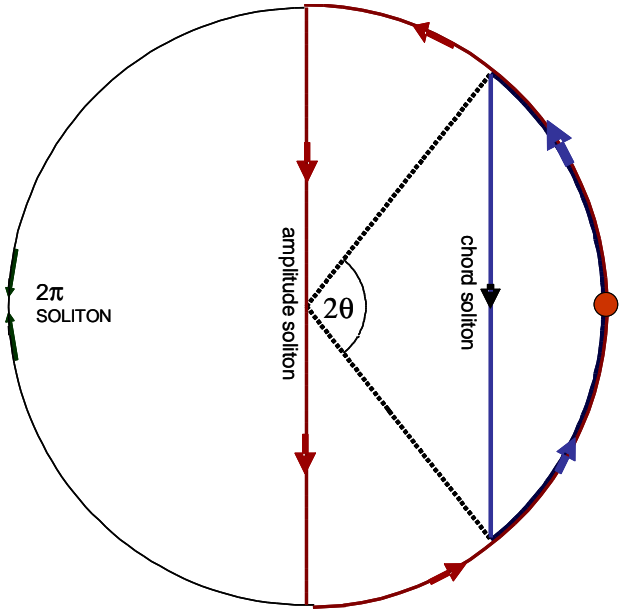}
\end{center}
\caption{Soliton trajectories in the complex plane of the order parameter. The circle is the manifold of allowed ground states $|\Delta|=\Delta_{0}$. The vertical diameter line $2\theta=\pi$ is the stable amplitude soliton. The vertical chordus line is an intermediate soliton within a chiral angle $2\theta$. The value $2\theta=100^{\circ}$ is chosen, which corresponds to the optimal configuration for the interchain tunneling (see \cite{brazov:03} and below). The arcs with arrows show phase tails required to level out the perturbation at large distances from the soliton.}
 \label{fig:chord-sol}
\end{figure}

The order parameter of the ICDW is the complex field $\Delta=A(x,t)\exp[i\varphi(x,t)]$, acting upon electrons by mixing states near the Fermi momenta points $\pm k_{F}$. Solitons in the ICDW rise from strong interactions between electronic $e,h$ and collective (amplitude $A$ and phase $\varphi$) degrees of freedom, which is set up since all of them are modified by spontaneous formation of the new GS \cite{brazov:89}. Electrons' dressing by the collective deformations results in relatively heavy self-trapped particles with lower energies $\Delta<\Delta_{0}$, which originates two basic scales. The high one can be observed, as $2\Delta_{0}$ or $\Delta_{0}$, for instantaneous processes like optics, tunneling, photoemission when the ground state has no time to adapt itself to a perturbation. The lower activation $\Delta$ can be observed for fully relaxed dressed particles, which might dominate stationary effects of thermodynamics, kinetics, NMR, etc., see refs. in \cite{matveenko:05-ecrys}. The GS degeneracy gives a special character to these complex elementary excitations: they become \emph{topological solitons} \cite{brazov:89}. These are the trajectories connecting different equivalent GSs; they are characterized by a chiral angle $2\theta=\varphi(x=+\infty)-\varphi(x=-\infty)$. This angle defines a family of "chordus solitons" - Fig.\ref{fig:chord-sol}, see Appendix for details; it provides a continuous path in the configurational space for electron's self-trapping. The time evolution $\theta(t)$ describes the self-trapping dynamics - the instanton \cite{matveenko:05-ecrys,brazov:03}, see Fig.\ref{fig:time} below.

\begin{figure}[tbh]
\begin{center}
\includegraphics[width=0.5\textwidth]{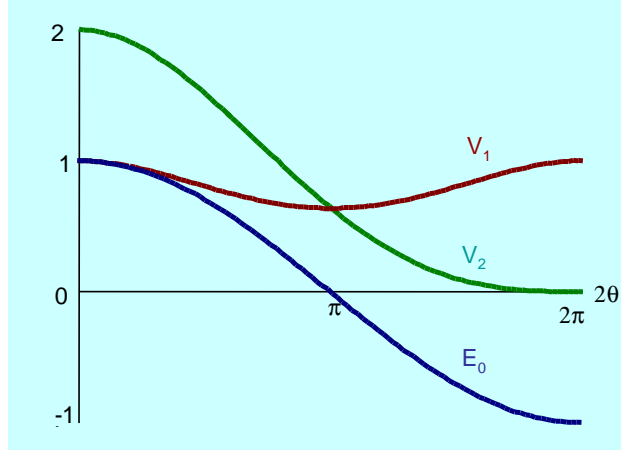}
\end{center}
\caption{Selftrapping branches (in units of $\Delta_{0}$) of chordus solitons:
total energies $V_{\nu}(\theta)$ for midgap fillings $\nu=1,2$ of localized split-off state and its energy level $E_{0}=\Delta_{0}\cos(\theta)$ as functions of $\theta$.}
 \label{fig:theta-branches}
\end{figure}

Processes of self-trapping of electrons or their pairs are important in several respects. They provide:
 \newline1. Spectral flow by transferring the split-off state $E_{0}$ between the gap edges $\Delta_{0}\rightarrow-\Delta_{0}$, Fig.\ref{fig:theta-branches};
  \newline2. Particle flow (conversion of normal carriers to the collective GS) between the conduction and the valence bands, which crosses the gap riding upon the split-off state;
   \newline3. Microscopic Phase Slips, by adding/subtracting the $2\pi$ winding of the order parameter.

\begin{figure}[tbh]
\begin{center}
\includegraphics[width=0.5\textwidth]{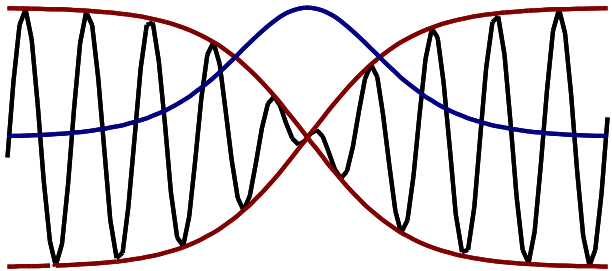}
\end{center}
\caption{Profiles for the amplitude soliton: oscillating electronic density, its overlap, and the spin density distribution due to the unpaired electron at the midgap state.}
 \label{fig:ampl-sol}
\end{figure}

For single-particle processes (see reviews \cite{brazov:89,braz-84}) the self-trapping of one electron at $\Delta_{0}$ or of one hole at $-\Delta_{0}$ will proceed gaining the energy until the configuration takes a stable form of the amplitude soliton AS: $\Delta(x)=\Delta_{0}\tanh(x/\xi_{0})\cos(Qx+\varphi)$, Fig.\ref{fig:ampl-sol}. Now $E_{0}=0$ becomes a pure midgap state occupied by a single electron; thus the soliton carries the electronic spin $s=1/2$. Curiously, the electric charge is zero rather than $e$, being compensated by the dilatation of electronic wave functions of the filled band $E<-\Delta_{0}$. It can be interpreted in a way that the AS is symmetric with respect to the charge conjugation: it is the adaptation of an electron added to the GS of 2M particles, as well as of a hole upon the GS of 2M+2 particles. (Breaking of charge conjugation can be still treated within exact solutions \cite{bm:84} - it gives a small electric charge to the AS.) Thus, in a 1D system, the AS is a realization of a spinon, the particle carrying the elementary spin 1/2 but no charge. This particle plays a central role in phenomenological pictures of strongly interacting electronic systems in general \cite{pwa:98}. Theoretical value for the AS energy is $W_{as}=2/\pi\Delta_{0}\approx0.65\Delta_{0}$; thus, the energy $\approx0.35\Delta_{0}$ is gained by converting the electron into the soliton.

Excitation of the solitonic state takes place in those moments, when collective quantum fluctuations create an appropriate configuration with the necessary split-off intragap state \cite{matveenko:05-ecrys}. At first sight, one needs to prepare the AS in its full form (the diameter in the Fig.\ref{fig:chord-sol}), the probability of which is very low. But actually the spontaneous deformation is more shallow. Indeed, the energy $E_{s}$ is yielded by the transfer of a single electron; hence the split-off energy level, prepared for this electron by the optimal fluctuations, must be at $E_{0}$~= $W_{as}$; therefore, the tunneling takes place when quantum fluctuations accumulate the chiral angle such that $\cos\theta$~= $2/\pi$, which gives $2\theta\approx100^{\circ}$, rather than $180^{\circ}$.

The tunneling threshold $2W_{as}$ of the pair-breaking is required only for the single particle channel, while for the bi-particle channel there may be no need for an activation as if the system is still metallic. This expectation resides upon the fact that the GS of the ICDW can incorporate a finite even amount of particles without paying an activation energy, unlike for odd numbers. For a finite length, the energy cost will be the same as for the metal: for each accommodated pair it is $\sim\hbar v_{F}/L$ - vanishing for the large system length $L$. The accommodation requires for the total CDW phase increment along the chain $\Delta\varphi=2\pi$ per the added pair (see the branch $V_{2}$ at the Fig.\ref{fig:theta-branches}). For the zero energy charge exchange between chains, it is necessary to adjust the number of states to the number of particles: that is at the chain $a$ to transfer the emptied level at $-\Delta_{0}$ just below the gap across the gap up to the manyfold of properly empty levels above $+\Delta_{0}$; the same time at the chain $b$ the empty level at $+\Delta_{0}$ must be drawn down across the gap to form a new level at $-\Delta_{0}$ which will accommodate the arrived pair. This job is done by a simultaneous phase slip processes of opposite directions: negative $0\Rightarrow-2\pi$ at the chain $a$, and positive $0\Rightarrow2\pi$ at the chain $b$. With the vanishing, in 1D regime, of the phase slip energy $W_{ps}\rightarrow 0$, the optimal fluctuation will require for the exactly mid-gap split-off state $E_0\rightarrow 0$, i.e. for $\theta\rightarrow \pi/2$, which is the AS. Hence the AS becomes an optimal configuration, the barrier, for nucleation of the phase slip.

\textbf{Effects of 2D,3S long range order.}
In systems of correlated chains the local configuration must conform to the long range order. At long distances the phase $\varphi$ must return to the mean value ($\varphi_{\infty}=0$ in our choice of Fig.\ref{fig:ampl-sol}).

The pure AS (vertical diameter) is accompanied by phase tails (arc lines) adjusting the phase discontinuity $\pi/2\rightarrow-\pi/2$ to the asymptotic zero value. With the total increment $2\times\pi/2$~= $\pi$ of the phase concentrated within these tails, they carry the electric charge $e$, in addition to the spin 1/2 concentrated within the AS core. The composed particle carries both of the electron quantum numbers, but they are localized at very different scales. Contrary to the spin, the charge is not associated to a particular electron level; the electric charge transfer by the tails of the AS is rather the collective conduction promoted by ASs. The price for the adaptation to the long range order is an increase of the AS energy by an amount of the order of the transition temperature $T_c$. For the phase soliton it becomes the scale of its total energy $W_{ps}\sim T_c$.

\subsection{Electrons and solitons in coherent tunneling experiments}

Recent tunneling experiments \cite{latyshev:05-ecrys} performed on quasi-1D materials with CDWs, namely NbSe$_{3}$ and TaS$_{3}$, can be interpreted as the first direct observation of solitons in dynamics. Tunneling spectra of NbSe$_{3}$ \cite{zandt,latyshev:05-ecrys} show sharp peaks which are identified with the intergap tunneling at $2\Delta_{2}$ and $2\Delta_{1}$ for the two CDWs coexisting in this material (see figures 1b and 2 in \cite{latyshev:05-ecrys}). Additional subgap features appear for both CDWs, when the zero bias contribution of normal electrons is suppressed by high magnetic field or by temperature T (see figure 2 in \cite{latyshev:05-ecrys}). These quite strong peaks scale together with $\Delta(T)$ at $V_{as}\approx2/3\Delta$. This unexpected feature finds a natural interpretation in the above picture of solitons as special elementary excitations. $V_{as}$ appears as the threshold for the tunneling between the Fermi level (middle of the gap) states of the electronic pocket, specific to $\mathrm{NbSe_3}$, and the nonlinear multi-electronic complex - the Amplitude Soliton.

The tunneling conductivity is also observed at even much lower voltages: $U>V_{t}$ (see figures 1a,3a,5a in \cite{latyshev:05-ecrys}), which can be provided only by the bi-electronic channel \cite{brazov:ecrys-05,matveenko:05-ecrys}. In the ICDW it is realized via $2\pi$ phase solitons, which correspond to one period of stretching/squeezing of one chain with respect to the surrounding ones. These are the elementary particles with the charge $\pm 2e$ and the energy $W_{ps}\sim T_{c}$ (recall their recent visualization in \cite{brun}), which gives the right order of magnitude both for the deep subgap tunneling range and for the longitudinal conductivity activation. The tunneling process corresponds to a coincidence of oppositely directed phase slips at neighboring chains, which transfer a pair of particle between their GS. A theory of these processes will be presented below in the section \ref{bi-e}.

\section{ Self-trapping processes.}
\subsection{Techniques.}

We shall follow the adiabatic method \cite{matv-ecrys02,matveenko:05-ecrys,brazov:03,mb:02,mb:05}, developed for the X-ray absorption (PES, ARPES), optics, and for tunneling in low symmetry systems supporting polarons. It assumes a smallness of collective frequencies, here of $2K_{F}$ phonons $\omega_{ph}$, in comparison with the electronic gap: $\omega_{ph}\ll\Delta_{0}$. Now, electrons are moving in a slowly varying potential $\Delta(x,t)$, so that at any instance $t$ their energies $E_{j}(t)$ and wave functions $\Psi_{j}(x,t)$ are defined from a stationary Schroedinger equation $H\Psi(x,t)=E(t)\Psi(x,E(t))$. The Hamiltonian $H=H(x,\Delta(x,t))$ depends on the instantaneous configuration $\Delta(x,t)$, so that $E(t)$ and $\Psi(x,E(t))$ depend on time only parametrically. The time will be chosen along the imaginary axis where the necessary saddle points of the action are commonly believed to be found (see e.g. \cite{LO}). (Examples of profound analysis of the complex time contours for typical self-trapping problems can be found in the review \cite{IR-review}.)

The process is determined by the change $\nu$, $|\nu|=0,1,2$, of the number of electrons; it is a relative filling factor of the split-off intragap state ($\nu>0$ for an electron and $\nu<0$ for a hole).
At a given $\nu$, the instantaneous configuration $\Delta(x,t)$ determines the action $S(\nu)\equiv S(\nu,\Delta)$. In the imaginary time, this functional is given as a sum of the kinetic and the potential $V_{\nu}$ energies:
\begin{equation}
S(\nu,\Delta)=\int dt\left\{\int dx
 \frac{\left\vert \partial_{t}\Delta\right\vert^{2}}
 {g_{ep}^{2}\omega_{ph}^{2}}+V_{\nu}(\Delta(x,t))\right\}
\,;\;V_{\nu}=V_{0}+|\nu|E
 \label{L(nu)}
\end{equation}
where $g_{ep}$ is the electron-phonon coupling constant and $\omega_{ph}$ is the bare phonon frequency. The potential $V_{\nu}$ contains the energy of deformations $g_{ep}^{-2}\int dx|\Delta|^{2}$, and the sum over electron energies in filled states (counted with respect to the GS energy for the total number $2M$ particles). It includes both the filled vacuum states at $E<-\Delta_{0}$ and the split-off intragap ones $E_{0}$, $-\Delta_{0}\leq E_{0}\leq\Delta_{0}$.  In the non perturbed, $\Delta\equiv\Delta_{0}$, state $V_{\nu}=|\nu|\Delta_{0}$: the particle added to the non deformed GS, is placed at the lowest allowed energy, $\Delta_{0}$. More details are given in the Appendix.

The self-trapping evolution of the added electron can be fortunately described by the known (see \cite{braz-84}) exact solution for intermediate configurations characterized by the singe intragap state $E_{0}=\Delta_{0}\cos\theta$ with $0\leq\theta\leq\pi$. It is known to be the Chordus Soliton with $2\theta$ being the total chiral angle: $\Delta(+\infty)/\Delta(-\infty)=\exp(2i\theta)$, see Fig.\ref{fig:theta-branches} and details in the Appendix.

In calculations we follow the approximation \cite{mb:02,matveenko:05-ecrys,mb:05,brazov:03} of the zero dimensional reduction
\begin{equation}
\Delta_{j}(x,t)\Rightarrow\Delta_{sol}(x-X_{j}(t),\theta_{j}(t))~;~
 S(\Delta_{j}(x,t))\Rightarrow S(\theta_{j}(t),X_{j}(t))
  \label{proj}
\end{equation}
which limits the whole manyfold of functions $\Delta_{j}(x,t)$ to a particular class of a given function $\Delta_{sol}$ of $x$ (relative to a time dependent center of mass coordinate $X_{j}$ and the local phase $\varphi_{j}$). The function $\Delta_{sol}$ is chosen as the chordus soliton trajectory which gives just one and only one split-off state $E_{0}=\Delta_{0}\cos\theta$.

The total on-chain action becomes
\begin{equation}
S=S(t_{1},t_{2},U)=S_{core}+S_{dis}
\label{S-total}
\end{equation}
were $S_{dis}$ is the dissipative contribution (\ref{S_snd}) described below and in the Appendix, while
$S_{core}$ is the local action from the core of the chordus soliton
\begin{equation}
S_{core}=
 \int_{t_{1}}^{t_{2}}dt\left\{f(\theta)\dot{\theta}^{2}+V_{\nu}(\theta)\right\}
 \label{S_core}
\end{equation}
Here $2f(\theta)$ is the variable "effective mass" of the "$\theta$-particle":

\begin{equation}
f(\theta)=\int dx\,\frac{1}{g_{ep}^{2}\omega_{ph}^{2}}\left\vert \frac
{d\Delta_{s}}{d\theta}\right\vert^{2}\label{f}
\end{equation}
and $V_{\nu}(\theta)$ is the potential branch (see Appendix and Fig.\ref{fig:theta-branches}). Its basic form is
\begin{equation}
V_{\nu}(\theta)=\Delta_{0}[(|\nu|-\frac{2}{\pi}\theta)\cos\theta
 +\frac{2}{\pi}\sin\theta]
  \label{V-nu}
\end{equation}

As the topologically nontrivial object, the chordus soliton cannot be created in a pure form with a finite chiral angle. In general, adaptational deformations must appear to compensate for the topological charge. These deformations are developing over long space-time scales and they can be described in terms of the gapless mode, the phase $\varphi$, alone. Hence, allowing for the time evolution of the chiral angle $\theta\rightarrow\theta(t)$ within the core, we should also unhinder the field $\varphi \rightarrow\varphi(x,t)$ at all $x$ and $t$. Starting from $x\rightarrow -\infty$ and finishing at $x\rightarrow\infty$, the system follows closely the circle $|\Delta|=\Delta_{0}$, changing almost entirely by phase - notice the arc lines at the Fig.\ref{fig:chord-sol}. Approaching the soliton core centered at $x=X_{s}$, the phase $\varphi(x\rightarrow X_{s}\pm0,t)$ matches the angles $\pm\theta$, which delimit the chordus part of the trajectory, see Fig.\ref{fig:chord-sol}. From large scales we view only a jump $\varphi(x\rightarrow X_s,t)\approx\theta(t)\mathrm{sgn}(x-X_{s})$, which effect can be easily extracted if we generalize the scheme suggested earlier for static solitons at presence of interchain coupling \cite{matv-ecrys02,soliton3d}. The action for the phase mode is
\begin{equation}
S_{snd}[\varphi(x,t),\theta(t)]=\frac{v_{F}}{4\pi}\iint
dxdt\left((\partial_{t}\varphi/u)^{2}+(\partial_{x}\varphi-2\theta(t)\delta(x-X_{s}))^{2}\right)
\label{S-phi}
\end{equation}
where $u$ is the CDW phase velocity:
$u=g_{ph}\omega_{ph}/\sqrt{4\pi}$.
 Integrating out
$\varphi(x,t)$ from $\exp\{-S_{snd}[\varphi,\theta]\}$ at a given
$\theta(t)$, we arrive at the typical action for the problem of
quantum dissipation \cite{leggett}:

\begin{equation}
S_{dis}[\theta]\approx\frac{v_{F}/u}{2\pi^{2}}\int\int
dtdt^{\prime}\left(
\frac{\theta(t)-\theta(t^{\prime})}{t-t^{\prime}}\right)^{2}=
 -\frac{v_{F}/u}{2\pi^{2}}\int\int dtdt^{\prime}
  \dot{\theta}(t)\ln|t-t^{\prime}|\dot{\theta}(t^{\prime})
   \label{S_snd}%
\end{equation}
that is $S_{dis}\sim\sum|\omega||\theta_{\omega}|^{2}$. (The above equivalence of two forms for $S_{dis}[\theta]$ holds for closed trajectories of $\theta(t)$.) The dissipation comes from emission of phase phonons, while forming the long range tail in the course of the chordus soliton development. In real time, the dissipation comes from emitting the phase fronts $\varphi(t,x)\approx Sgn(x)\theta(t-|x|/u)$.

\section{One- and two- particle tunneling processes}
Consider the system of two chains ($j=a,b$) weakly coupled via hybridization of their electronic states, which amplitude is $t_{\bot}\ll\Delta_{0}$. The chains are maintained at the electric potential difference $U>0$. Electrons jump between the chain $a$ at the potential $U/2$ and the chain $b$ at the potential $-U/2$. Initially both chains are in their ground states with the same total number of particles $2M$ at one chain. In principle, this is a nonequilibrium situation with the total energy difference of $2MU$. By definition, we do not allow for an extensive $\delta M\sim M$ interchain charge transfer to form a common Fermi level: that would be an inhomogeneously reconstructed junction state \cite{latyshev-prl:06}, which requires for a special treatment. Considering rare events of one or two particle tunneling, we need to follow only the potential difference for their intragap states.

For the interchain tunneling Hamiltonian
\begin{equation}
H_{\perp}=\sum_{<a,b>}t_{\perp}\int
dx(\Psi_{a}^{\dag}(x,t)\Psi_{b}(x,t)
+\Psi_{b}^{\dag}(x,t)\Psi_{a}(x,t)) \label{Hab}%
\end{equation}
the transverse electric current operator is
\begin{equation}
{\hat{\jmath}}_{\perp}=it_{\perp}(\Psi_{a}^{\dag}(x,t)\Psi_{b}(x,t)
 -\Psi_{b}^{\dag}(x,t)\Psi_{a}(x,t)). \label{oper}%
\end{equation}
The average current (number of electrons per chain per unit length per unit time) is given by the trace of this operator over fermions and the functional integral over lattice deformations.
\begin{equation}
j(U)=\int\int D\Delta_{a}(x,t)D\Delta_{b}(x,t)Tr\hat{\jmath}_{\perp}
\exp\left[-S(\Delta_{a})-S(\Delta_{b})-\int dt H_{\perp}\right] \label{jU}
\end{equation}
We shall need $j(U)$ only in the lowest nonvanishing order of expansion over $t_{\perp}$. Each application of operators $j_{\bot},H_{\bot}$ at a time $t$ produces following effects: changing the occupation numbers of split-off states $\nu_{j}$ by $\pm1$, and multiplication by the product of corresponding wave functions $\Psi_{a}^{\ast}(x,t)\Psi_{b}(x,t)$, adding the voltage energy gain $U$.

\subsection{One electron tunneling.\label{one-e}}

Consider the single particle channel in interchain tunneling for a standard ICDW, with a double $g=2$ spin degeneracy of electronic states.

Processes originated by the transfer of one electron between the chains appear already in the lowest second order of expansion of the current in powers of $t_{\perp}$. The contribution $J_{1}$ to the interchain current can be written as a functional integral over fields
$\Delta_{j}$, $j=a,b$:
\begin{multline*}
J_{1}\sim t_{\perp}^{2}\int d(x-y)\int d(t_{1}-t_{2})\int
D[\Delta_{j}]
 [\Psi_{a}^{\ast}(x,t_{1})\Psi_{b}(x,t_{1})\Psi_{a}(y,t_{2})\Psi_{b}^{\ast}(y,t_{2})\\
\exp(-S_{1}(t_{1},t_{2},\Delta_{j}(x,t)))
\end{multline*}
were $\psi_j$ are wave functions of the particle added and extracted in
moments $t_{1},t_{2}$ at fluctuational intragap levels $E_j$. Here the time
dependent action $S(t_{1},t_{2})$ describes the process of transferring one
particle from the doubly occupied level $E_{a}<0$ of the chains $a$ to the
unoccupied level $E_{b}>0$ of the chain $b$ at the time $t_{1}$, and the
inverse process at the time $t_{2}$. We have

\begin{eqnarray}
S_{1}(t_{1},t_{2},\Delta_{j}(x,t)) &  =S_{a}(0,-\infty,t_{1})+S_{b}
(0,-\infty,t_{1})+S_{a}(0,t_{2},\infty)+S_{b}(0,t_{2},\infty)\\
&  +S_{a}(-1,t_{1},t_{2})+S_{b}(1,t_{1},t_{2})+U(t_{1}-t_{2})
\label{S1}
\end{eqnarray}
where $S_{j}(\nu,t_1,t_{2})=S_{j}(\nu,t_1,t_{2};\Delta_j)$ are the on-chain actions (\ref{S-total}).
The branches' components are shown in the table below ( $\theta$ stays for $\theta^{a}$ or for $\pi-\theta^{b}$), signs $\pm$ correspond to $a$ and $b$):

\begin{table*}[h]
\begin{tabular}{|c|c|c|}
\hline
$E/\Delta_{0}$ & $V_{0}/\Delta_{0}\text{(}t<t_{1},t>t_{2}\text{)}$
 &
 $V_{1}/\Delta_{0}\text{ (}t_{1}<t<t_{2}\text{)}$\\[5pt]
 \hline
$-\cos\theta\pm U/2~ $&$ -\frac{2}{\pi}\theta\cos\theta+\frac{2}{\pi}\sin\theta$
 &
$(1-\frac{2}{\pi}\theta)\cos\theta+\frac{2}{\pi}\sin\theta-U/2$ \\[5pt]
\hline
\end{tabular}
\end{table*}

\begin{figure}[tbh]
\begin{center}
\includegraphics[width=8cm]{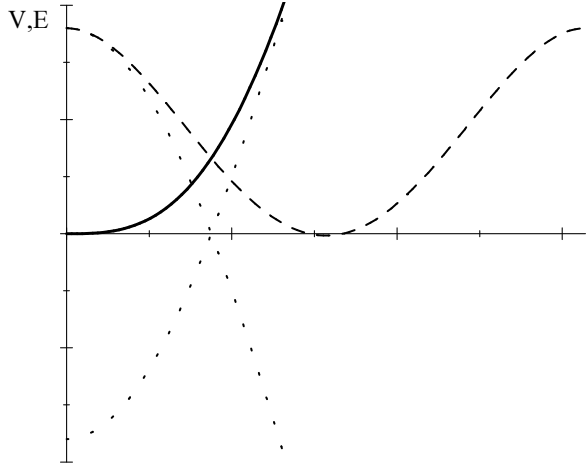}
\end{center}
\caption{Branches for the one electron tunneling. The potential evolves following the ascending solid line of $V_0$, then switching to the descending dashed line of $V_1$ until it touches the axe $V=0$ Energy levels are shown as dotted lines.}%
\label{fig:one-e}%
\end{figure}
The corresponding branches are shown at the Fig.\ref{fig:one-e}. At this plot we have chosen a critical, minimal value of $U=2W_{s}=\allowbreak1.27\Delta_{0}$ (the thick line touched the zero level), that is $\theta_{i}=\arccos2/\pi\approx50^{\circ}$.

The main contribution to the subgap current comes from extremal instanton trajectories, which minimize the action $S_{1}$ (\ref{S1}) over both $\Delta_{j}(x,t)$ and $\tau=t_{2}-t_{1}$, at a given $U$, thus yielding $\min{S}=S(U)$. The exact extremal trajectory is defined by equations
\begin{equation}
\delta S/\delta\Delta_{j}(x,t)=0~;\
 \partial S/\partial t_{1}=0~,~
 \partial S/\partial t_{2}=0.
 \label{dS}%
\end{equation}

The first equation in (\ref{dS}) becomes, in our projection, $\delta S/\delta\theta_{j}(x,t)=0$, hence the same equation for $\theta_{a}$ and $\pi-\theta_{b}$:
\begin{equation}
-\frac{d}{dt}\left[f(\theta)\left(\frac{d\theta}{dt}\right)^{2}\right]
+\dot{\theta}\frac{\partial}{\partial\theta}V(\theta,t)
 +\frac{v_{F}}{\pi^{2}u}\dot{\theta}(t)\int_{-\infty}^{\infty}dt^{\prime}
 \frac{\theta(t)-\theta(t^{\prime})}{(t-t^{\prime})^{2}}=0
  \label{deq0}%
\end{equation}
The equation (\ref{deq0}) describes the dissipative motion of an effective $\theta-$ particle with the variable mass $2f$ in the effective potential $-V_{\nu(t)}(\theta)$, which depends explicitly on $t$ via jumps of $\nu=\nu(t)$ at the tunneling moments $t_{i}$.

The second pair of equations in (\ref{dS}) can be written, in our model, as
\begin{equation}
{E_{0}}_{a}(t_{i})+{E_{0}}_{b}(t_{i})=U~~;\ E_{0}(t_{i})
 =\Delta_{0}\cos\theta(t_{i})=\Delta_{0}\cos\theta_{i}
 \label{E-U}%
\end{equation}
This equation shows that the tunneling takes place when quantum fluctuations at both chains simultaneously create such potential wells, that they produce split-off intragap states satisfying the resonance condition $E_{b}-E_{a}=U$. (The last statement is literally valid only within our model neglecting Coulomb interactions, see examples of these corrections in \cite{matveenko:05-ecrys,mb:05}.)

Based on earlier results \cite{matv-ecrys02,brazov:03} augmented by reasonable assumptions (see Appendix), we conclude that, up to pre-exponential factors, the tunneling current is proportional to the square of the single particle spectral density, which has been already calculated \cite{matv-ecrys02,brazov:03}, as the PES intensity $I_{PES}$: $J_{1}\propto t_{\perp}^{2}I_{PES}^{2}(\Omega=U/2)$. E.g. near the threshold $U\approx2W_{as}$
\begin{equation}
J_{1}(U)=At_{\perp}^{2}\left(\frac{U-2W_{as}}{W_{as}}\right)^{2\beta}
 \,,\;\beta=\frac{v_{F}}{4u}\gg1
  \label{J1low}%
\end{equation}
This law gives the vanishing current at the threshold, which is different from low symmetry cases \cite{mb:02,minton:92} where the cutoff was finite. A physical origin of this suppression comes from efficient emittance of gapless phonons in the course of tunneling, which drives dynamics to the regime of quantum dissipation \cite{matv-ecrys02,brazov:03}.

Calculating the prefactor $A$ requires for integration over $\Delta_{j}(x,t)$ around the extremal, taking into account the zero modes related with translations of the instanton centers positions $X_{j}(t)$ and with the phases $\varphi_{j}=\varphi(X_{j})$. The prefactor $A=A(U)$ in (\ref{J1low}) is a slow function of $U$; it may yield powers of $(U-2W_{as})^{-\gamma}$ with $\gamma\sim 1 \ll\beta$ which dependence is negligible in comparison with the big power $2\beta$. But the $U$ dependence of the prefactor becomes important in the 3D regime near the threshold. The 3D long range ordering, which energy scale is measured by the transition temperature $T_{c}\ll\Delta_{0}$, returns the tunneling to the normal dynamics. Then the peak of $J_{1}(U)$, instead of zero (\ref{J1low}), will develop in a narrow vicinity $0<U-2W_{s}<T_{c}$ of the threshold, in a qualitative agreement with experiment \cite{latyshev-prl:05}. The factor $\left({{T_{c}}/W_{s}}\right)^{v/2u}$ will give the overall reduction of the current in comparison with the one expected for free electrons.

Behavior near the free edge $\Omega\approx\Delta_{0}$ is governed by small fluctuations $\eta$ of the gap amplitude $|\Delta |=\Delta_{0}+\eta$ and of the Fermi level $\delta E_{F}=\varphi^{\prime}v_{F}/2$ via the phase gradient $\varphi^{\prime}=\partial_{x}\varphi$. The gap fluctuations dominate over the phase ones, so the results are similar to the ones for commensurate CDWs. A thorough discussion was given in \cite{brazov:03}.

\subsection{Spinless case.}
\begin{figure}[tbh]
\begin{center}
\includegraphics[width=0.5\textwidth]{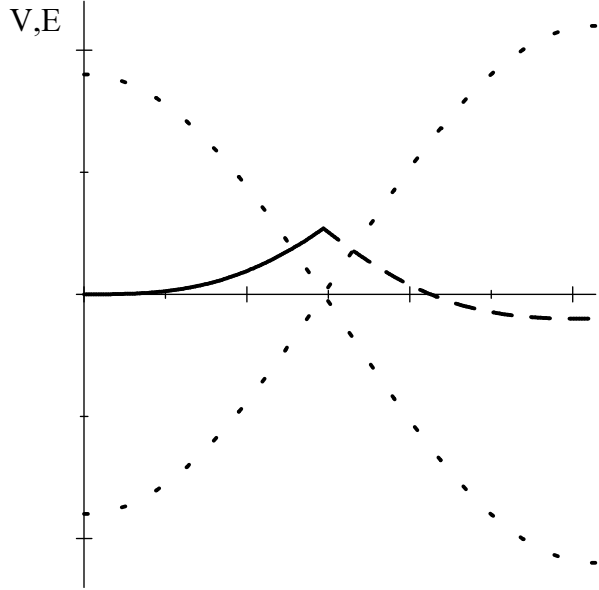}
\end{center}
\caption{Branches for the spinless case. Here $U=0.2$, then $\theta_{i}=\arccos(0.1)=1.47rad$. Initial branch $V_{0}$ at $\theta<\theta_{i}$ is given by the solid line, the branch $V_{1}$ at $\theta>\theta_{i}$ - by the dashed line, energies $E_0^j$ are shown as doted lines.}%
\label{fig:spinless}%
\end{figure}

The spinless case, the degeneracy $g=1$, applies to spin polarized CDWs in high magnetic fields, which is a current experimental trend \cite{hmf}. It describes also the XY model of the spin-Peierls state in magnetic chains. Also this case provides a tutorial to the regime of full phase slips at small $U$, which otherwise appear only in higher forth order in $t_{\perp}$ (see the next section). The interchain jump of the fermion takes place between the filled level rising from $E=-\Delta_0$ and the empty level descending from $E=\Delta_0$. The branches' components are shown in the following table and drawn at the Fig.\ref{fig:spinless} (all energies are in units of $\Delta_0$).

\begin{table*}[h]
\begin{tabular}{|c|c|c|}
\hline
$E$ & $V_{0}\text{ (}t<t_{1},t>t_{2}\text{)}$
 &
$V_{1}\text{ (}t_{1}<t<t_{2}\text{)}$\\[5pt]
 \hline
$ -\cos\theta\pm U/2$
 &
$ -\frac{1}{\pi}\theta\cos\theta+\frac{1}{\pi}\sin\theta $&
$(1-\frac{1}{\pi}\theta)\cos\theta+\frac{1}{\pi}\sin\theta-U/2$\\[5pt]
\hline
\end{tabular}
\end{table*}

At $U\rightarrow0$ the tunneling point moves to $\theta_{i}\rightarrow\pi/2$, where the configuration is the
amplitude soliton. It provides the exactly midgap state, which is necessary at $U\rightarrow 0$, hence there is no low boundary for $U$. We arrive at the result
\begin{equation}
\tilde{J}_{1}\sim
t_{\perp}^{2}\left(\frac{U}{W_{as}}\right)^{\beta}\,,\;\beta=\frac{v_{F}}{4u}\gg1
\end{equation}
(The exponent is twice smaller in comparison with (\ref{J1low}) because
coefficients of each term in the action for $g=1$ are $1/2$ of those for $g=2$.)

\subsection{Bi-electronic tunneling.\label{bi-e}}

The joint self-trapping of two electrons allows to further gain the energy, resulting in stable states different from independent ASs: there are the phase slips leading to phase $2\pi$ solitons. An advantage of tunneling experiments is the possibility to see the phase slips directly, at voltages $U$ below the two AS threshold $U<2W_{as}$ of the pair-breaking, i.e. within the true single particle gap. The probability of the bi-electronic tunneling is relatively small as it appears only in the higher order $\sim t_{\perp}^{4}$ in interchain coupling. But it can be seen as extending below the one-electron threshold, where no other excitations can contribute to the tunneling current.

Consider the forth order ($\sim t_{\perp}^{4}$) contribution to the current due to the possibility of electron pairs tunneling between chains.

\begin{align}
J_{2} &  \propto t_{\perp}^{4}\int\prod_{j=a,b} D\Delta_{j}(x,t)
 \int\prod_{i=1}^{4}dy_{i}dt_{i}\exp(-S_{2})
 \nonumber\\
&  [\Psi_{b}^{\ast}(y_{1},t_{1})\Psi_{a}(y_{1},t_{1})
 \Psi_{b}^{\ast}(y_{2},t_{2})\Psi_{a}(y_{2},t_{2})
 \Psi_{a}^{\ast}(y_{3},t_{3})\Psi_{b}(y_{3},t_{3})
 \Psi_{a}^{\ast}(y_{4},t_{4})\Psi_{b}(y_{4},t_{4})
 \nonumber\\
& -\{y_{4},t_{4}  \longleftrightarrow y_{3},t_{3}\}]\hfill
 \label{t4}
\end{align}

\begin{eqnarray}
S_{2}=S_{2}(t_{1..4},\Delta_{j}(x,t))  =S_{a}(0,-\infty,t_{1})+
 S_{b}(0,-\infty,t_{1})+S_{a}(0,t_{4},\infty)+S_{b}(0,t_{4},\infty)
 \nonumber\\
+S_{a}(-1,t_{1},t_{2})+S_{b}(1,t_{1},t_{2})+S_{a}(-1,t_{3},t_{4})+S_{b}(1,t_{3},t_{4})
 -U(t_{2}-t_{1}+t_{4}-t_{3})
 \nonumber\\
+S_{a}(-2,t_{1},t_{2})+S_{b}(2,t_{1},t_{2})-2U(t_{2}-t_{1})
 \label{2e}
\end{eqnarray}
The action (\ref{2e}) describes the following sequence: The first jump takes place at $t=t_{1}$, $\theta(t_{1})=\theta_{1}$ from the twice filled level rising from $E=-\Delta_0$ to the empty level descending from $E=\Delta_0$. Its branches are the same as shown above in the table for the one-electron tunneling.

\begin{figure}[ptb]
\begin{center}
\includegraphics[width=0.5\textwidth]{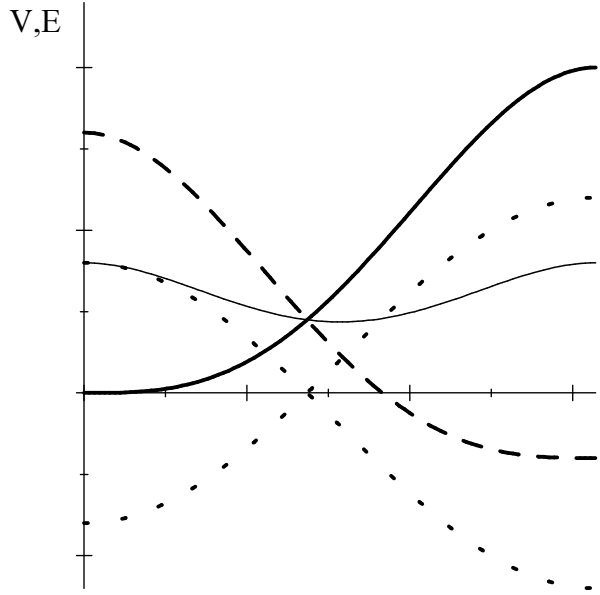}
\end{center}
\caption{Branches for the bi-electron tunneling. Here $U=0.4$, then the transition (lines crossing) takes place at
$\theta=\arccos(0.2)=1.37$ - below the amplitude soliton value $\theta=\pi/2$. Split-off energies are shown by dotted lines. The potential evolves following the thick lines: solid, then dashed ones. The transition of each of two electrons proceed via the intermediate branch $V_1$ - the thin line.}
\label{fig:bi-e}%
\end{figure}

The second jump takes place at $t=t_{2}$, $\theta(t_{2})=\theta_{2}$ from the singly filled level of the chain $a$ to the singly filled level of the chain $b$. The branches' components are shown in the following table and plotted at the Fig.\ref{fig:bi-e}. (All energies are in units of $\Delta_0$.)

\begin{table*}[h]
\begin{tabular}{|c|c|c|}
\hline
$V_{0}\text{ (}t<t_{1},t>t_{4}\text{)}$ &
$ V_{1}\text{ (}t_{1}<t<t_{2}\text{),(}t_{3}<t<t_{4}\text{)}$ &
$  V_{2}\text{ (}t_{2}<t<t_{3}\text{)}$\\[5pt]
 \hline
$-\frac{2}{\pi}\theta\cos\theta+\frac{2}{\pi}\sin\theta $ &
$(1-\frac{2}{\pi}\theta)\cos\theta+\frac{2}{\pi}\sin\theta-U/2 $&
$(2-\frac{2}{\pi}\theta)\cos\theta+\frac{2}{\pi}\sin\theta-U $\\[5pt]
\hline
\end{tabular}
\end{table*}

Rigorously, the extremal trajectory is defined as above in Sec.\ref{one-e} from equations
\[
\frac{\delta S}{\delta\theta_{j}(x,t)}=0~,~j=a,b\ ;\ \
\frac{\partial S}{\partial t_{i}}=0~,~i=1..4
\]
Similar considerations argue that equations are consistent provided
\begin{equation}
\theta_{a}(x,t)\equiv\pi-\theta_{b}(x,t)\ \ ;\
 E_{0}(t_{i})=\frac{U}{2};\ t_{1}=t_{2}~,~\,t_{3}=t_{4}\ \
  \label{a0}%
\end{equation}
The last condition $t_{1}=t_{2}~,~\,t_{3}=t_{4}$ tells that only processes of simultaneous tunneling of pairs of particles contribute to the extremal action (\ref{2e}). The branch with $\nu=1$ is strongly virtual, it is passed over the microscopic time $\xi_{0}/u$, hence at the Fig.\ref{fig:bi-e} the interval of $V_{1}$ is reduced to the point. The intermediate state of $\nu=1$ gives no contribution because of the accidental degeneracy of our model. This is the artifact of the Peierls model: due to the conditions (\ref{a0}) the curves $V_{0}$, $V_{1}-U/2$, $V-U$ intersect at the one point $\theta$ where $E_{0}=U/2$. At $U\rightarrow0$ the tunneling points move to $\theta_{1,2}\rightarrow\pi/2$, where the configuration is the amplitude soliton.

Finally we arrive at the power law with the index $v_{F}/u$ twice that for the single particle process
\begin{equation}
J_{2}\propto\exp[-2S]\propto
t_{\perp}^{4}\left(\frac{U}{\Delta_{0}}\right)
^{v_{F}/u}\label{i2}%
\end{equation}
The bi-electronic channel in ICDW stretches the PG down to the small energy $\propto T_{c}$, i.e. even to $U=0$ in the 1D limit. The reason is that here the particles with the charge $2e$ - the phase $2\pi$ solitons are neither bipolarons, nor pairs of kinks as for systems with lower symmetries, see \cite{brazov:ecrys-05,latyshev-prl:06}. The tunneling takes a form of a coincidence of opposite $\pm2\pi$ phase slips taking place simultaneously at adjacent chains. In the regime with the 2D or 3D long range order, the current $J_{2}$ will vanish below the low threshold $2W_{ps}$ where the characteristic intensity will be of the order of
 $J_{2}\propto t_{\perp}^{4}\left({T_c}/\Delta_0\right)^{v_F/u}$.

\begin{figure}[htb]
    \includegraphics[width=0.5\textwidth]{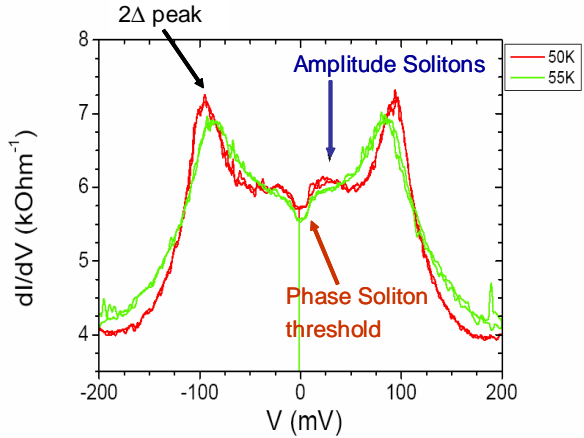}\\
  \caption{(Color on line.) Experimental data from the collection of Latyshev et al, e.g. \cite{latyshev:05-ecrys}, and their interpretation. The features correspond to the second, low $T$, CDW in $\mathrm{NbSe_3}$. They are taken just below the transition temperature ($T=55K$ - green curve), and at slightly lower $T=50K$ - red curve. The evolution of the AS feature from an edge to a peak corresponds to our guesses on the role of the long range order.}
  \label{fig:exp}
\end{figure}

\section{Conclusions}
We conclude that the three regimes are expected theoretically for the internal interchain tunneling in ICDWs.
\newline 1. High energy single-particle peak at $2\Delta_0$ corresponding to the electron-hole excitation across the rigid CDW gap.
\newline 2. Middle energy edge at $2W_{as}\approx 4/3\Delta_0$ corresponding to creation of two Amplitude Solitons, which are, unlike undressed electrons of the case \#1, the exact pair-breaking eigenstates of the whole system.
\newline 3. Low energy edge at $W_{ps}\sim T_c$, which drops down even to $U=0$ in the 1D regime, when the scale of the interchain ordering energy $T_c$ may be neglected.

The regimes \#1 and \#2 can be accessible also in optical absorption, while the regime \#3 is specific to tunneling. Some more studies are necessary to reach a quantitative comparison with experiment. There are effects of a finite temperature and of direct electronic interactions, calculating the prefactors, taking into account the inhomogeneous distribution of the electric potential over a junction. Nevertheless, the general classification, assignment of features, and their location are already well covered by the theory, cf. \cite{latyshev-prl:06,brazov:ecrys-05}. These relations of theory and experiment are illustrated at the Fig.\ref{fig:exp}

\textbf{Acknowledgements} The authors are grateful to Yu.I. Latyshev and P. Monceau for numerous discussions. A support is acknowledged to the INTAS grant 7972 and to the ANR program (the project BLAN07-3-192276). S.I.M. acknowledges the support of the LEA Physique Theorique et Matiere Condensee and of the PAST program, and the hospitality of the LPTMS.

\section{Appendix}

\subsection{Selftrapping details.}

In the following the units of energy and length are $\Delta_{0}\Rightarrow1$ and $\xi_{0}=\hbar v_{F}/\Delta_{0}\Rightarrow1$.

Topologically nontrivial trajectories in the plane of the complex order parameter $\Delta$ are conveniently parameterized by a family of chordus solitons

\begin{equation}
\Delta_{sol}=E_{0}-ik_{0}\tanh(k_{0}x)
\end{equation}
Here the split-off energy $E_{0}$ and the inverse localization length $k_{0}$ are parameterized by the chiral angle $2\theta$ (see Fig.\ref{fig:chord-sol}) as
\[
\ E_{0}=\pm\cos\theta~,\ k_{0}=\sin\theta~;\ 0\leq\theta\leq\pi
\]
with signs $\pm$ corresponding to two directions of the trajectory (the positive one is shown at Fig.\ref{fig:chord-sol}). We shall assume that $\pi>\theta>0$, so that the negative chirality is obtained by transformation $\theta\Rightarrow\pi-\theta$. The level $E_{0}$ is filled by $0\leq\nu\leq g$ fermions, while vacuum levels at $E<-\Delta_{0}$ are all filled with the total spin degeneracy factor $g$. (In applications, $g=2$ - electrons, or $g=1$ - spinless fermions.) The spectral flow provided by the positive phase slip draws the energy level $E_{0}=\cos\theta$ down from $E_{0}=1$ at $\theta=0$ towards $E_{0}=-1$ at $\theta=\pi$. Its energy branch is
\begin{equation}
V_{\nu}(\theta)=(\nu-\frac{g}{\pi}\theta)\cos\theta+\frac{g}{\pi}\sin
\theta~,\ \ E_{0}=\cos\theta~,\ \nu=0,...,g
 \label{V}%
\end{equation}
The spectral flow provided by a negative phase slip draws the energy level $E_{0}=-\cos\theta$ up from $E_{0}=-1$ at $\theta=0$ towards $E_{0}=1$ at $\theta=\pi$. Its energy branch is $V_{-\nu}(\theta)=V_{\nu}(\pi-\theta)$.

\subsection{Dissipative dynamics.}
Here we shall analyze general properties of the extremal solutions.
The total on-chain action can be written as
\begin{eqnarray}
S(\theta,\nu)=\int dt\left\{f(\theta)\dot{\theta}^{2}+V_{\nu}(\theta)\right\}
-\frac{v_{F}/u}{2\pi^{2}}\int\int
dtdt^{\prime}\dot{\theta}(t)\ln|t-t^{\prime}|\dot{\theta}(t^{\prime})
\label{S-tot}
\end{eqnarray}

Its variation, Eq.(\ref{deq0}), can be written as

\begin{equation}
2f\ddot{\theta}+f^{\prime}\dot{\theta}^{2}-V^{\prime}+\frac{v_{F}}{\pi^{2}u}
 \int_{-\infty}^{\infty}dt^{\prime}\frac{\dot{\theta}(t^{\prime})}{(t-t^{\prime})}=0
 \label{t-1}%
\end{equation}
Here $\dot{\theta}$ means the time derivative, while $^{\prime}$ means the
derivative over $\theta$.

\begin{figure}
\begin{center}
\includegraphics[width=0.5\textwidth]{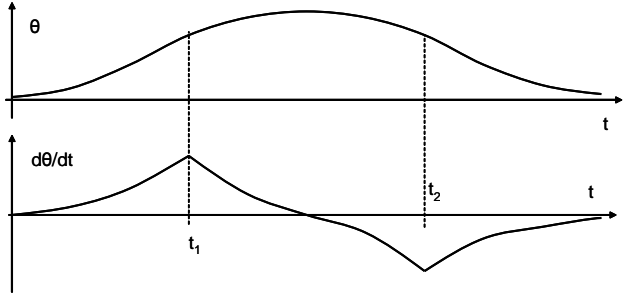}
\end{center}
\caption{Schematic plots of time dependencies for $\theta$ and $\dot{\theta}$ for one-electron tunneling.}\label{fig:time}
\end{figure}

Consider the single-particle process when $\nu$ jumps from $0$ to $\pm1$ and back at times $t_{1,2}$. The expected time dependencies for $\theta$ and $\dot{\theta}$ are shown schematically at the Fig.\ref{fig:time}.

Let $\tau=t_2-t_1 \rightarrow\infty$, then the solution should decouple in two solitons evolving between $0\Rightarrow\theta_{0}$ and $\theta_{0}\Rightarrow0$, and we may consider only the the first one. Here $\theta_{0}$ is the stable exited state corresponding to the minimum of $V_{1}$ : $\theta_{0}=\pi/2$ for the spin 1/2 case and $\theta_{0}=\pi$ for the spinless case. Contrary to usual problems with degenerate vacuums, here there is a discontinuity at the floating point $t_{1}$ when $0<\theta _{1}<\theta_{0}$. At the level $\theta_{1}$, $V^{\prime}$ has a jump, hence the jump in $\ddot{\theta}$ and the cusp in $\dot{\theta}$.  The result will confirm that the total increment of $\theta$: $\Delta\theta=\theta_{0}$ is accumulated in a finite vicinity of $t_{1}$, then Eq.(\ref{t-1}) becomes
\begin{equation}
2f\ddot{\theta}+f^{\prime}\dot{\theta}^{2}-V^{\prime}\approx
 -\frac{v_{F}}{\pi^{2}u}\frac{\theta_{0}}{t-t_{1}}
 \label{asym}
\end{equation}
Here $\dot{\theta}$ means the time derivative, while $^{\prime}$ means the derivative over $\theta$. At the left asymptotics $t-t_{1}\rightarrow-\infty$, $V=V_{0}\sim\theta^{3}g/3\pi$ and $f\sim\theta$, then the l.h.s. of (\ref{asym}) is dominated by the term $V^{\prime}$ and we obtain the asymptotics $\theta\sim\left\vert t-t_{1}\right\vert^{-1/2}$. The right asymptotics at $t-t_{1}\rightarrow\infty$ depends on the degeneracy. For $g=1$, the final state $\theta_{0}=\pi$ is the full phase slip, equivalent to the initial state $\theta=0$; hence the right asymptotics is the same as the left one. For $g=2$, $\theta_{0}=\pi/2$ corresponds to the amplitude soliton. Here $f\rightarrow cnst$ and $V_{1}\approx\allowbreak\frac{2}{\pi}+\frac{1}{\pi}\left(\theta-\frac{1}{2}\pi\right)^{2}$; again the l.h.s. of (\ref{asym}) is dominated by the term $V^{\prime}$ and we obtain the asymptotics $\theta-\theta_0 \sim\left\vert t-t_{1}\right\vert^{-1}$, and similar near $t_2$. The obtained asymptotics show that the integral in the dissipative term of Eq.(\ref{t-1}) is convergent around the impact points $t_i$, which makes our assumptions to be consistent.

For a finite $\tau$ we have a two-soliton process, and, in the leading order the solution is the sum of two contributions. Then for the outer interval $|t|\gg \tau$, $\theta\sim \tau t^{-3/2}$. Within the inner interval $t_{1}<t<t_{2}$
\[
\theta_{0}-\theta\sim\left\{\left(t-t_{1}\right)^{-1}+
 \left(t_{2}-t\right)^{-1}\right\}
  \sim\frac{\tau}{(\tau/2)^{2}-(\Delta t)^{2}}~,\ \Delta t=t-\frac{t_{2}+t_{1}}{2}
\]
We find for the turning point $\theta_{mx}-\theta_{0}\sim1/\tau$.

The same procedure can be applied to calculating the action (\ref{S-tot}): again, the integral is concentrated at $t,t^{\prime}$ being near the points $t_{1},t_{2}$. When both $t,t^{\prime}$ belong to the same vicinity, we obtain a constant contribution to the action which may be omitted. When  $t$ and $t^{\prime}$ belong to vicinities of different points $t_{1}$ or $t_{2}$, then the contribution is $\sim\ln \tau$. Collecting only the contributions which increase with $\tau$, we obtain the action $S\tau$, its extremum $S_{extr}=S(\tau_{extr})$, and finally the tunneling probability $J_{1}\sim\exp(-2S_{extr})$:

\begin{align*}
S  &
\approx(V_{1}(\theta_{_{0}})-U)\tau+\frac{v_{F}}{u}
 \left(\frac{\theta_{0}}{\pi}\right)^{2}\ln\tau~;\
  \tau_{extr}\approx\frac{v_{F}}{u}
  \left(\frac{\theta_{0}}{\pi}\right)^{2}\frac{1}{(U-U_{min})}~;\\
\ J_{1}  & \sim\left(  \frac{U-U_{min}}{U_{min}}\right)^{2\beta} ~,\
\beta=\frac{v_{F}}{u}\left(\frac{\theta_{0}}{\pi}\right)^{2} \ ,\
U_{min}=V_{1}(\theta_{_{0}})
\end{align*}

Consider finally the question of uniqueness of the chosen solution.
Phases $\theta_{j}$ obey (up to the opposite rotation $\theta\Rightarrow\pi-\theta$) same equations, characterized by common switching moments $t_{i}$. Their solutions could be different in principle, as characterized by different sets of switching phases $\theta_{i}^{j}=\theta^{j}(t_{i})$, $i=1,2$ and $j=a,b$. Eqs. (\ref{E-U}) reduce twice the number of boundary variables, so that e.g. only $\theta_{i}^{a}$ are independent unknowns. The instanton process starts at $t=-\infty$ with values $\theta^{j}=0$, comes to the moment $t_{1}$ at values $\theta_{1}^{a}$ and $\theta_{1}^{b}$ related by (\ref{E-U}), reaches the turning points $\theta_{mx}^{j}$ at turning moments $t_{mx}^{j}$; then it comes to the moment $t_{2}$ with values $\theta_{2}^{a}$ and $\theta _{2}^{b}$ related by (\ref{E-U}), and finally returns to values $\theta^{j}=0$ at $t=+\infty$, see Fig.\ref{fig:time}. We shall assume that the process is symmetric, as it is demonstrated by solutions of nondissipative models, see \cite{matveenko:05-ecrys,mb:05}, and of simpler purely dissipative ones \cite{LO,korshunov:87a,korshunov:87b}. Then $\theta_{2}^{j}=\theta_{1}^{j}$, and $t_{mx}^{j}=(t_{1}+t_{2})/2$ becomes the same for both chains. We are left with a single unknown boundary phase at each chain, uniquely related to time interval $\tau=t_{2}-t_{1}$. The parameter is common for both chains, then the solutions $\theta^{j}(t)$ are identical. In view of (\ref{E-U}), they will be determined by the same condition ${E_{0}}_{a}(t_{i})={E_{0}}_{b}(t_{i})=U/2$. We arrive at the effective one chain problem with a doubled effective action for the current.

\end{document}